\documentclass[12pt]{iopart}

\usepackage{graphicx}

\begin{document}

\title[Exploring phase transitions of a multi-qubit--cavity system]
      {Exploring super-radiant phase transitions via coherent control of a 
       multi-qubit--cavity system}

\author{Timothy C Jarrett, Chiu Fan Lee and Neil F Johnson}

\address{Centre for Quantum Computation and Physics Department, 
  University of Oxford, Clarendon Laboratory, Parks Road, Oxford OX1 3PU, U.K.}
  
\eads{\mailto{t.jarrett1@physics.ox.ac.uk}, \mailto{c.lee1@physics.ox.ac.uk}
  \mailto{n.johnson@physics.ox.ac.uk}}

\begin{abstract}
We propose the use of coherent control of a multi-qubit--cavity QED system in 
order to explore  novel phase transition phenomena in a
general class of multi-qubit--cavity systems. In addition to atomic systems, the
associated super-radiant phase transitions should be observable in a variety of
solid-state experimental systems, including the technologically important case of
interacting quantum dots coupled to an optical cavity mode.
\end{abstract}

\submitto{\JOB}


\section{Introduction}
There is much current interest in the use of coherent control in order to 
generate novel matter-radiation states in cavity QED and atom-optics systems 
\cite{Alexandra}. In addition, the field of cavity QED has caught the interest of 
workers in the field of solid-state nanostructures, since effective two-level 
systems can be fabricated using semiconductor quantum dots, organic molecules and 
even naturally-occuring biological systems such as the photosynthetic complexes 
LHI and LHII and in biological imaging setups involving FRET (Fluoresence 
Resonance Energy Transfer) \cite{Photo}. Such nanostructure systems could then be 
embedded in optical cavities or their equivalent, such as in the gap of a 
photonic band-gap material \cite{pbg}. We refer to Ref. \cite{Alexandra2} for a 
discussion of the size and energy-gaps of the artificial nanostructure systems 
which can currently be fabricated experimentally.

In a parallel development, phase transitions in quantum systems are currently 
attracting much attention within the solid-state, atomic  and quantum information 
communities \cite{Kad,Sac99,qip,nielsen}. Most of the focus within the 
solid-state community has been on
phase transitions in electronic systems such as low-dimensional magnets
\cite{Kad,Sac99} while in atomic physics there has been much interest in phase
transitions in cold atom gases and in atoms coupled to a cavity. In particular, a
second-order phase transition, from normal to superradiant, is known to arise in
the Dicke model which considers
$N$ two-state atoms (i.e. `spins' or `qubits' \cite{qip,nielsen}) coupled to an
electromagnetic field (i.e. bosonic cavity mode) 
\cite{Dic54,HL73,WH73}. The Dicke model itself has been studied within the atomic
physics community for fifty years, but has recently caught the attention of
solid-state physicists working on arrays of quantum dots, Josephson
junctions, and magnetoplasmas
\cite{CMP}. Its extension to quantum
chaos \cite{chaos}, quantum information
\cite{entanglement} and other exactly solvable models has also been
considered recently \cite{solvable}.  It has also been conjectured that 
superradiance could be used as a mechanism for quantum teleportation
\cite{teleportation}.

Here we extend our discussion in Ref. \cite{LJ04} on the exploration of novel phase 
transitions in atom-radiation systems exploiting the current levels of 
experimental expertise in the area of coherent control. The corresponding 
experimental set-up can be a cavity-QED, atom-optics, or nanostructure-optics 
system, whose energy gaps and interactions are tailored to be the required 
generalization of the well-known Dicke model \cite{WH73}. We show that, according 
to the values of these control parameters, the phase transitions be driven to 
become first-order.

\section{The Model}
The well-known Dicke model from atom-optics ignores interactions between the 
constituent two-level systems or 
`spins' \cite{WH73}. In atomic systems
where each `spin' is an atom, this is arguably an acceptable approximation if the
atoms are neutral {\em and} the atom-atom separation
$d\gg a$ where $a$ is the atomic diameter. However there are several reasons why
this approximation is unlikely to be valid in typical solid-state systems. First,
the `spin' can be represented by any nanostructure (e.g. quantum dot) possessing 
two
well-defined energy levels, yet such nanostructures are not typically neutral. 
Hence
there will in general be a short-ranged (due to screening) electrostatic 
interaction
between neighbouring nanostructures.  Second, even if each nanostructure
is neutral, the typical separation distance
$d$ between fabricated and self-organised nanostructures is typically the same as
the size of the nanostructure itself. Hence neutral systems such as excitonic
quantum dots will still have a significant interaction between nearest neighbors
\cite{us}.

Motivated by the experimental relevance of `spin--spin' interactions, we 
introduce
and analyze a generalised Dicke Hamiltonian which is relevant to current
experimental setups in both the solid-state and atomic communities \cite{expt}. 
We
show that the presence of transverse spin--spin coupling terms, leads to novel
first-order phase transitions associated with super-radiance in the bosonic 
cavity
field. A technologically important example within the solid-state community would
be an array of quantum dots coupled to an optical mode. This mode could arise 
from an
optical cavity, or a defect mode in a photonic band gap material \cite{expt}.
However we emphasise that the $N$ `spins' may correspond to  {\em any} two-level
system, including superconducting qubits and atoms \cite{CMP,expt}. The bosonic
field is then any field to which the corresponding spins couple \cite{CMP,expt}.
Apart from the experimental prediction of novel phase transitions, our work also
provides an interesting generalisation of the well-known Dicke model.

The method of solution that we present here is in fact valid for a wider 
class of Hamiltonians incorporating spin--spin and spin--boson interactions
\cite{unpub}.  We follow the method of Wang and Hioe \cite{WH73}, whose results
also proved to be valid for a wider class of Dicke Hamiltonians.
We focus on the simple example of the Dicke Hamiltonian with an additional
spin--spin interaction in the $y$ direction.
\begin{eqnarray} H&=&a^\dag a +
 \sum_{j=1}^{N} \left\{ \frac{\lambda}{2 \sqrt{N}}
 (a +a^\dag)(\sigma^+_j+  
 \sigma^-_j) + \frac{\epsilon}{2} \sigma^Z_j  - J \sigma^Y_j \cdot
\sigma^Y_{j+1} \right\} \\ &=&a^\dag a +
 \sum_{j=1}^{N} \left\{\frac{\lambda}{\sqrt{N}}
 (a +a^\dag)\sigma^X_j + \frac{\epsilon}{2} \sigma^Z_j 
 - J \sigma^Y_j \cdot \sigma^Y_{j+1} \right\} \ .
\end{eqnarray}
Following the discussion above, the experimental spin--spin interactions are 
likely
to be short-ranged and hence only nearest-neighbor interactions are included in 
$H$. 
The operators in Eqs. 1 and 2 have their usual, standard meanings.

\section{Results}
To obtain the thermodynamical properties of the system, we first introduce the
Glauber coherent states  $|\alpha\rangle$ of the field \cite{Glauber}
where 
$a|\alpha \rangle = \alpha | \alpha \rangle$, 
 $\langle \alpha | a^\dag = \langle \alpha | \alpha^*$.
The coherent states are complete,
$\int \frac{d^2\alpha}{\pi}  |\alpha \rangle
\langle \alpha| =1$.
In this basis, we may write the canonical partition function as:
\begin{equation} 
Z(N,T)=\sum_{\bf s} \int 
\frac{d^2\alpha}{\pi} 
\langle {\bf s} | \langle \alpha| e^{-\beta H} | \alpha \rangle | {\bf s}
\rangle
\end{equation}
As in Ref. \cite{WH73}, we adopt the following assumptions:
\begin{enumerate}
\item
$a/\sqrt{N}$ and $a^\dag/ \sqrt{N}$ exist as $N \rightarrow \infty$;
\item
$\lim_{N \rightarrow
\infty} \lim_{R \rightarrow \infty} \sum_{r=0}^R \frac{(-\beta H_N)^r}{r!}$ can
be interchanged
\end{enumerate}
We then find 
\begin{equation}
Z(N,T) = \int \frac{d^2 \alpha}{\pi} e^{-\beta|\alpha|^2}{\rm Tr} e^{-\beta H'}
\end{equation} where 
\begin{equation} 
H' = \sum_{j=1}^N \left\{\frac{2 \lambda {\rm Re}( \alpha)}{\sqrt{N}} \sigma^X_j
 + \frac{\epsilon}{2} \sigma^Z_j  -J \sigma^Y_j \cdot \sigma^Y_{j+1} \right\}.
\end{equation}
We first rotate about the $y$-axis to give
\begin{equation}
H' = -J\sum_{j=1}^N
 \left\{ \sqrt{\left( \frac{2 \lambda {\rm Re}( \alpha)}{J\sqrt{N}}\right)^2
+\left( \frac{\epsilon}{2J}\right)^2} \sigma^{Z'}_j 
 +   \sigma^Y_j \cdot \sigma^Y_{j+1} \right\}.
\end{equation}
We note here that the resulting hamiltonian is of the type of an Ising 
hamiltonian with a transverse field, and it exhibits a divergence in concurrence 
at its quantum phase transition (see, e.g., \cite{qip}). This particular model is 
instrumental in understanding the nature of coherence in quantum systems. Going 
back to the calculations, we may now diagonalise $H'$ by performing a 
Jordan-Wigner transformation,
passing into momentum-space and then performing a Bogoliubov transformation
(see, for example, Ref. \cite{Sac99}).  We then have, in terms of momentum-space
fermion operators $\gamma_k$, the diagonalised $H'$:
\begin{equation}
\label{epsilon1}
H' = \sum_{k=1}^N \xi_k(\alpha) (\gamma_k^\dag \gamma_k -\frac{1}{2} )
\end{equation}
with
\begin{eqnarray}
\label{epsilon2}
\xi_k (\alpha) &=& 2J \sqrt{1+(g(\alpha))^2 - 2g(\alpha) } \\ 
g(\alpha) &=&
\sqrt{\left( \frac{2 \lambda {\rm Re} ( \alpha)}{J\sqrt{N}}\right)^2
	+\left( \frac{\epsilon}{2J}\right)^2 } \ .
\end{eqnarray}
We may then write
\begin{equation}
H' = \sum_{k=1}^N H_k
\end{equation}
where
\begin{equation}
H_k = \xi_k(\alpha) (\gamma_k^\dag \gamma_k -\frac{1}{2} ).
\end{equation}
From the transformation, we may associate the spin-up state with an empty orbit
on the site and a spin-down state with an occupied orbital.  Using the
commutation relations for the $\gamma_k$ and the fact that $\gamma_k|0\rangle =
0$ (see, for example, Ref. \cite{Sac99}), we obtain
\begin{equation}
Z(N,T) = \int \frac{d^2 \alpha}{\pi} e^{-\beta |\alpha|^2} 
\prod_{k=1}^N\{e^{-\frac{\beta}{2}\xi_k(\alpha)}+e^{\frac{\beta}{2}\xi_k(\alpha)}
\}.
\end{equation}
Writing $d^2\alpha = d{\rm Re}(\alpha)d{\rm Im}(\alpha)$, $w = {\rm
Re}(\alpha)$ and integrating out ${\rm Im}(\alpha)$ we obtain
\begin{equation}
Z(N,T) = \frac{1}{\sqrt{\beta}{\pi}} \int dw e^{-\beta w^2 + \sum_{k=1}^N
  \left\{ \log \left[ \cosh \left(\frac{\beta}{2} \xi_k(x)  \right) \right]
    + \log (2) \right\}} \ .
\end{equation}
We now let $x=w/\sqrt{N}$. Writing $\sum_{k=1}^N$ as $\frac{N}{2 \pi}
\int_0^{2\pi} dk$,
yields \begin{equation} 
Z(N,T)=\sqrt{\frac{N}{\beta \pi}}
\int_{-\infty}^\infty dx \left\{ e^{- \beta x^2 + I(x)} \right\}^N
\end{equation} where
\begin{eqnarray} I(x)&=& \frac{1}{2 \pi} 
\int_0^{2 \pi} dk \left\{ 
\log \left[ \cosh \left(
\frac{\beta}{2} \xi_k(x)  \right) \right] + \log (2) \right\} 
\end{eqnarray} and
\begin{equation}
\xi_k (x) = 2J \sqrt{1+(g(x))^2 - 2 g(x) \cos k } \ .
\end{equation} 
From here on, we omit the $\log (2)$ term in $I(x)$ since
it only contributes an overall factor to $Z(N,T)$. 

Laplace's method now tells
us that
\begin{equation}
\label{laplace1} Z(N,T) \propto \max_{-\infty \leq x \leq \infty} \exp \left\{
N [-\beta x^2 + I(x)] \right\}.
\end{equation}
Denoting $[-\beta x^2 + I(x)] $ by $\Omega(x)$, we recall that the
super-radiant phase corresponds to $\Omega(x)$ having its maximum at a non-zero
$x$ \cite{WH73}. If there is no transverse field, i.e., if $J=0$, and the
temperature is fixed, then the maximum of $\Omega(x)$ will split continuously
into two maxima symmetric about the origin as $\lambda^2$ increases.  Hence the 
process is a
continuous phase transition.  

\begin{figure}
\caption{Demonstration as to what the function $\Omega(x)$ looks like across a 
first-order phase transition as $\lambda$ and $x$ are varied. Here $J=1.0$, 
$\epsilon=1.1$ and $\beta=100.0$.}
\begin{center}
\includegraphics{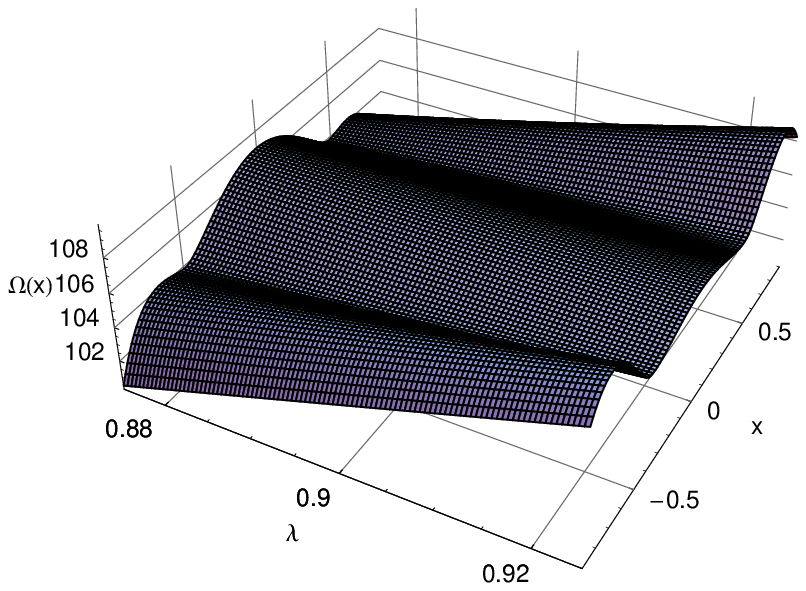}
\end{center}
\label{PT}
\end{figure}

However the case of non-zero $J$ is qualitatively different from $J=0$. As a 
result
of the frustration induced by the tranverse nearest-neighbour couplings, {\it 
there
are regions where the super-radiant phase transition becomes first-order.} Hence 
{\em the system's phase transition can be driven to become first-order by 
suitable adjustment of the nearest-neighbour couplings}.
This phenomenon of first-order phase transitions is
revealed by considering the functional shape of $\Omega(x)$, as shown in Fig. 
\ref{PT}.

\begin{figure}
\caption{The value of $x$ at which there is a maximum in $\Omega(x)$ as 
 $\lambda$ increases, for $J=1.0$ (dashed line) and $J=0.8$ (solid line). In both 
cases 
$\epsilon=1.1$ and $\beta=100$.}
\begin{center}
\includegraphics{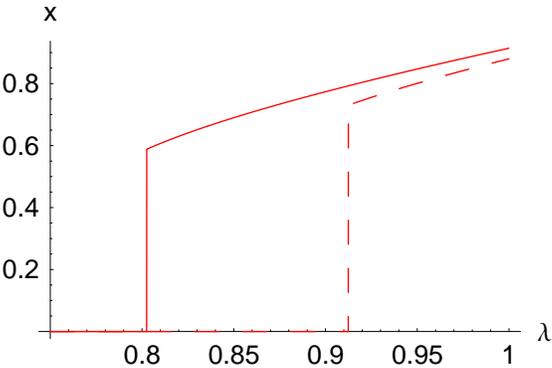}
\end{center}
\label{x-lambda}
\end{figure}

Figure \ref{x-lambda} shows the value of $x$ that maximises $\Omega(x)$ at fixed 
$\epsilon$
and two different values of $J$.  From the two lines, we can see that the
spin--spin coupling actually acts to inhibit the phase transition.  As we increase $J$ from
$0.8$ to $1.0$ we can see that the value to which we have to increase $\lambda$
to induce a phase transition is higher.

\begin{figure}
\caption{The maximiser of $\Omega$ shown as a  function of
$J$ and $\epsilon$. 
Here $\lambda=1.3$ and $\beta=100$.}
\begin{center}
\includegraphics{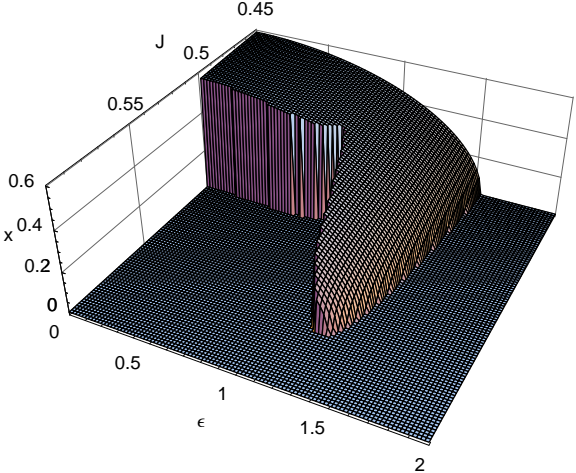}
\end{center}
\label{Maximiser}
\end{figure}

Figure \ref{Maximiser} plots the maximiser of $\Omega(x)$ with $\lambda$ fixed at
a value of 1.3.  
For small $J$, the local (non-zero) maximum of $\Omega(x)$ converges to 
zero as we increase $\epsilon$ and the system is no longer super-radiant.
This is no longer the case if $J$ is increased. In this case, $\Omega(x)$ has a
global maximum when $\epsilon$ is small; however as $\epsilon$ increases, the
non-zero local maxima becomes dominant and as a result a first-order phase 
transition occurs.  We note that the barriers between the wells are infinite in
the thermodynamic limit, hence we expect that the sub-radiant state is
metastable as $\epsilon$ increases. This observation also 
suggests the phenomenon of hysteresis, which awaits experimental validation.

\begin{figure}
\caption{Plot of the order parameter, $\Theta = \langle \frac{a^\dag a}{N}
  \rangle$, for the phase transition with $\lambda=0.9$, $J=1.0$ and
  $\epsilon=1.1$.}
\begin{center}
\includegraphics{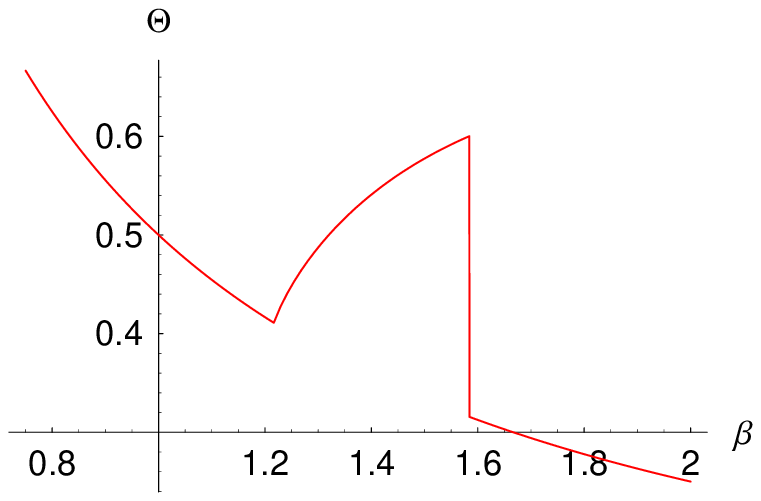}
\end{center}
\label{OP}
\end{figure}

In Fig. \ref{OP} we consider the order parameter of the transition,
$\langle\frac{a^\dag a}{N}\rangle$.  Following the same method as above, we
may calculate this to be equivalent to $x^2$ with an additional $\frac{1}{2 
\beta}$ term that comes from the imaginary part of the coherent states of the 
radiation field \cite{Duncan}.  We can see from the figure that as we lower 
$\beta$ we drive the system first through a first order phase transition and
then through a continuous phase transition.  Thus we are able to achieve both a
first and second order phase transition by varying the one parameter, $\beta$.

\section{Conclusion}
In conclusion, we have shown that the experimentally relevant spin-spin
interaction in the Dicke model transforms it into an Ising-hamiltonian with a 
photon-field dependent transverse field, which allows for an existence of both 
first-order and second-order phase transitions as parameters vary. 
Our
results highlight the importance of spin-spin coupling terms in spin-boson 
systems
and opens up the possibility of coherently controlling the competition
between the sub-radiant and super-radiant states in experimental atom-radiation 
systems \cite{unpub}.

\section*{References}
\begin{thebibliography}{99}

\bibitem{Alexandra} See, for example, A. Olaya-Castro, N. F. Johnson and L 
Quiroga, Phys. Rev. A {\bf 70}, 020301(R) (2004); J. Opt. B: Quantum Semiclass. 
Opt. {\bf 6}, S730 (2004); Phys. Rev. Lett., in press (2005).

\bibitem{Photo} A.J. Berglund, A.C.
Doherty and H. Mabuchi, Phys. Rev. Lett. {\bf 89}, 068101 (2002); J. Gilmore and 
R.H. McKenzie, cond-mat/0401444; A. Olaya-Castro, C.F. Lee and N.F. Johnson, in 
preparation (2005).

\bibitem{pbg} P.M. Hui and N.F. Johnson, {\em Solid State Physics}, Vol. 49, 
edited by H. Ehrenreich and F. Spaepen (Academic Press, New York, 1995).

\bibitem{Alexandra2} A. Olaya-Castro and N. F. Johnson, {\em Handbook of 
Theoretical and Computational Nanotechnology}, in press (2005); quant-ph/0406133.

\bibitem{Kad} L.P. Kadanoff, \emph{Statistical Physics} (World Scientific,
Singapore, 2000).

\bibitem{Sac99} S. Sachdev, \emph{Quantum Phase Transitions} (Cambridge
University Press, Cambridge, 1999).

\bibitem{qip} T.J. Osborne and M.A. Nielsen, Phys. Rev. A {\bf 66}, 032110 
(2002); 
R. Somma, G. Ortiz, H. Barnum, E. Knill, and L. Viola, quant-ph/0403035.

\bibitem{nielsen} M.A. Nielsen and I.L. Chuang, 
{\it Quantum Computation and Quantum
Information} (Cambridge  University Press, Cambridge, 2002).

\bibitem{Dic54} R.H. Dicke, Phys. Rev. {\bf 170}, 379 (1954).

\bibitem{HL73} K. Hepp and E.H. Lieb, Ann. Phys. (N.Y.) {\bf 76}, 360 (1973).

\bibitem{WH73} Y.K. Wang and F.T. Hioe, Phys. Rev. A {\bf 7}, 831 (1973); F.T.
Hioe, Phys. Rev. A {\bf 8}, 1440 (1973).

\bibitem{Glauber} R. Glauber, Phys. Rev. {\bf 131}, 2766 (1963).

\bibitem{CMP}
T. Vorrath and  T. Brandes, Phys. Rev. B {\bf 68}, 035309 (2003); 
W.A. Al-Saidi and  D. Stroud, Phys. Rev. B {\bf 65}, 224512 (2002); X. Zou,  K.
Pahlke and  W. Mathis, quant-ph/0201011; S. Raghavan,  H. Pu,  P. Meystre and 
N.P. Bigelow, cond-mat/0010140; N. Nayak,  A.S. Majumdar and  V. Bartzis, J.
Nonlinear Optics {\bf 24}, 319 (2000); T. Brandes,  J. Inoue and  A. Shimizu,
cond-mat/9908448 and cond-mat/9908447.

\bibitem{chaos} C. Emary and T. Brandes, Phys. Rev. Lett. {\bf 90}, 044101
(2003); Phys. Rev. E {\bf 67}, 066203 (2003).

\bibitem{entanglement} N. Lambert, C. Emary and T. Brandes, Phys. Rev. Lett.
{\bf 92}, 073602 (2004);  S. Schneider and G.J. Milburn, quant-ph/0112042; 
G. Ramon,  C. Brif and  A. Mann,
Phys. Rev. A {\bf 58}, 2506 (1998);  A. Messikh,  Z. Ficek and M.R.B.
Wahiddin, quant-ph/0303100.

\bibitem{solvable}
C. Emary and T. Brandes, quant-ph/0401029; 
S. Mancini,  P. Tombesi and V.I. Man'ko, quant-ph/9806034.

\bibitem{teleportation} Y.N. Chen et al., cond-mat/0502412. 

\bibitem{LJ04} C.F. Lee and N.F. Johnson, Phys. Rev. Lett. {\bf  93}, 083001 
(2004).

\bibitem{us} L. Quiroga and N.F.
Johnson, Phys. Rev. Lett. {\bf 83}, 2270 (1999); J.H. Reina, L. Quiroga and N.F.
Johnson, Phys. Rev. A {\bf 62}, 012305 (2000).

\bibitem{expt} E. Hagley et al., Phys. Rev. Lett. {\bf 79}, 1 (1997); A.
Rauschenbeutel et al., Science {\bf 288}, 2024 (2000); 
A. Imamoglu et al., Phys. Rev. Lett. {\bf 83}, 4204 (1999);  S.M. Dutra, P.L. 
Knight and H.
Moya-Cessa, Phys. Rev. A {\bf 49}, 1993 (1994); Y. Yamamoto and R. Slusher,
Physics Today, June (1993), p. 66; D.K. Young, L. Zhang, D.D. Awschalom and E.L.
Hu, Phys. Rev. B {\bf 66}, 081307 (2002); G.S. Solomon, M. Pelton and Y.
Yamamoto, Phys. Rev. Lett. {\bf 86}, 3903 (2001); B. Moller, M.V. Artemyev and U.
Woggon, Appl. Phys. Lett. {\bf 80}, 3253 (2002); N. F. Johnson, J. Phys. Condens. 
Matter 7, 965
(1995). 

\bibitem{unpub} C.F. Lee, T.C. Jarrett and N.F. Johnson, unpublished.

\bibitem{Duncan} G.C. Duncan, Phys. Rev. A {\bf 9}, 418 (1974).

\end{thebibliography}
\end{document}